\documentclass[onecolumn,nofootinbib,notitlepage,pra,showpacs,superscriptaddress,amssymb,amsmath,amsmath]{revtex4-1}
\usepackage{graphicx}
\usepackage{epstopdf}
\usepackage{bm}
\usepackage{hyperref}
\usepackage{comment}
\usepackage{color}
\usepackage{physics}
\usepackage{amsmath}
\usepackage{tikz}
\usepackage{enumitem}
\usepackage{bbold}
\usepackage{dsfont}
\usepackage[normalem]{ulem}

\hypersetup{%
   pdfpagemode=UseNone, 
   pdfstartpage=1,
   pdfmenubar=true,
   pdftoolbar=true,
   colorlinks = true,
   linkcolor=blue,
   citecolor=blue,
   urlcolor=blue,
   bookmarksopen=false}

\definecolor{fuchsia}{rgb}{1.0, 0.0, 1.0}
\definecolor{maroon}{rgb}{0.788, 0.0, 0.086}
\definecolor{ao}{rgb}{0.0, 0.5, 0.0}

\newcommand*{\la}{\langle}
\newcommand*{\ra}{\rangle}
\newcommand*{\btheta}{\boldsymbol{\theta}}
\newcommand*{\btildetheta}{\boldsymbol{\tilde{\theta}}}

\begin{document}

\title{Hessian-based toolbox for reliable and interpretable machine learning in physics}

\author{Anna Dawid}
\affiliation{Faculty of Physics,  University of Warsaw, Pasteura 5, 02-093 Warsaw, Poland}
\affiliation{ICFO - Institut de Ci\`encies Fot\`oniques, The Barcelona Institute of Science and Technology, Av. Carl Friedrich Gauss 3, 08860 Castelldefels (Barcelona), Spain}

\author{Patrick Huembeli}
\affiliation{Institute of Physics, \'Ecole Polytechnique F\'ed\'erale de Lausanne (EPFL), CH-1015 Lausanne, Switzerland}

\author{Micha\l~Tomza}
\affiliation{Faculty of Physics,  University of Warsaw, Pasteura 5, 02-093 Warsaw, Poland}

\author{Maciej Lewenstein}
\affiliation{ICFO - Institut de Ci\`encies Fot\`oniques, The Barcelona Institute of Science and Technology, Av. Carl Friedrich Gauss 3, 08860 Castelldefels (Barcelona), Spain}
\affiliation{ICREA, Pg.~Llu\'is Campanys 23, 08010 Barcelona, Spain}

\author{Alexandre Dauphin}
\affiliation{ICFO - Institut de Ci\`encies Fot\`oniques, The Barcelona Institute of Science and Technology, Av. Carl Friedrich Gauss 3, 08860 Castelldefels (Barcelona), Spain}

\date{\today}

\begin{abstract}
Machine learning (ML) techniques applied to quantum many-body physics have emerged as a new research field. While the numerical power of this approach is undeniable, the most expressive ML algorithms, such as neural networks, are black boxes: The user does neither know the logic behind the model predictions nor the uncertainty of the model predictions.
In this work, we present a toolbox for interpretability and reliability, agnostic of the model architecture. In particular, it provides a notion of the influence of the input data on the prediction at a given test point, an estimation of the uncertainty of the model predictions, and an extrapolation score for the model predictions. Such a toolbox only requires a single computation of the Hessian of the training loss function. Our work opens the road to the systematic use of interpretability and reliability methods in ML applied to physics and, more generally, science.
\end{abstract}


\maketitle


\section{Introduction}

One of the main challenges of quantum many-body physics is the identification of quantum phases of matter. Usually, physicists follow a traditional approach: with the help of physical intuition and educated guessing, they determine an order parameter governing transitions. Recently, an alternative route has been explored: machine learning (ML) algorithms can locate phase transitions, even in systems with highly non-trivial order parameters~\cite{Carrasquilla17NatPhys, Nieuwenburg17NatPhys}. Since then, deep fully connected and convolutional neural networks (CNNs) have been applied to detect phase transitions in a variety of physical models, for classical \cite{Carrasquilla17NatPhys, Li18, Schafer19, Cole2020, Liu2021}, quantum \cite{Nieuwenburg17NatPhys, Wetzel17a, Liu18PRL, Chng18, Huembeli19, Kottmann2020PRL, Arnold2021, Broecker17, Theveniaut19, Dong18, Blucher2020}, and topological~\cite{Zhang18PRL, Tsai19, Huembeli18, Greplova19, Balabanov2021} phase transitions with supervised \cite{Carrasquilla17NatPhys, Broecker17, Theveniaut19, Dong18, Blucher2020, Li18, Zhang18PRL, Tsai19, Cole2020} and unsupervised \cite{Huembeli18, Schafer19, Nieuwenburg17NatPhys, Wetzel17a, Liu18PRL, Chng18, Greplova19, Huembeli19, Kottmann2020PRL, Arnold2021, Balabanov2021, Liu2021} approaches as well as for experimental data \cite{Rem19, Khatami20, Kaming2021}.
Other examples include ML models that do not leverage deep architectures \cite{Wang16, Vargas18b}. However, the development of tools to build ML systems has generally outpaced the growth and adoption of tools to understand what they learn (\textit{interpretability} methods) and whether we can trust their predictions (\textit{reliability} methods).\\


\textbf{Interpretability.} The lack of interpretability is now a widely recognized challenge in the computer science community~\cite{Lipton2018, Carvalho19MDPI, Murdoch19PNAS, Molnar19book, Du20ACM} especially when ML is applied to real-world problems like medical diagnosis, insurance cost estimation, etc. 
In science, the lack of interpretability can be disturbing because the black-box behavior of the models prevents us from learning anything about novel physics.
Physicists are already addressing the need for interpretation of ML models, but the majority of proposed methods is either restricted to linear and kernel models~\cite{Wang16, Wetzel17b, Ponte17, Zhang19a, Greitemann19, Greitemann2021, Cole2020, Liu2021} or to the particular model architecture~\cite{Wetzel17b, Wetzel20, Arnold2021} or requires pre-engineering of the data, which limits the results nonuniversally to a
specific ML and physical model~\cite{Zhang19b, Balabanov2021}. An interesting, largely model-agnostic method in the context of lattice quantum field theory was proposed in~\cite{Blucher2020}.\\

\textbf{Reliability.} Another desired feature of ML models, which is intertwined with interpretability, is their reliability. A reliable ML model informs a user if its decisions are uncertain or result from pure extrapolation~\cite{Abdar2021}. In computer science, the reliability is especially important in the context of safety-critical problems or adversarial attacks, i.e., careful perturbations of input samples that aims to mislead the ML model on the test set~\cite{Biggio2018}. 
While data sets of physical interest are not endangered by such intentional attacks, adversarial ML tells us that tiny noise in the input may completely derail the model prediction. Moreover, while the reliability of ML in everyday problems can often be controlled by humans checking the predictions, it is improbable for, e.g., unknown phase diagrams. 
Nevertheless, the reliability of ML is not yet properly addressed in physics. 
While Bayesian ML, i.e., the most popular approach providing the uncertainty of predictions, proved to be a promising direction in molecular dynamics \cite{Krems2019}, it is generally difficult and needs a specific model architecture.
Therefore, there is a great need for more model-agnostic tools to estimate the uncertainty of ML predictions.\\


\textbf{Goal of the paper.} To address the need for interpretability and reliability in the detection of phase transitions with ML methods, in this work, we apply four ML interpretability and reliability tools to the CNN trained in a phase classification problem.
To better understand what a ML model learns, we extract the concept of similarity between input data from a machine. As a result, we can find out what is the relation between data according to the ML model and deduce what features are important for the classification. To this end, we employ and compare influence functions \cite{Koh17, Koh19} and relative influence functions (RelatIFs) \cite{Barshan2020}. Moreover, we address the need for model-agnostic assessment of the uncertainty of model predictions. We present Resampling Uncertainty Estimation (RUE) \cite{Schulam2020}, which allows for generating analogues of error bars for ML model predictions. Finally, we apply a tool called Local Ensembles (LEs)~\cite{Madras2020}, which warns a user if a ML model makes predictions with a high level of extrapolation.
We present these methods on the neural network trained to detect the quantum phase transition
in the one-dimensional (1D) spinless Fermi-Hubbard model.
The four methods require a single calculation of the Hessian of the training loss. Together, they form a Hessian-based toolbox that can be applied to any ML model and any learning scheme that relies on the calculation of the test and training loss functions and therefore finds application outside of physics and the detection of phase transitions.\\

\textbf{Structure of the paper.} This paper is structured as follows. Section \ref{s:methods} describes the used interpretability and reliability ML methods along with the Hessian and the concept of similarity. Section \ref{s:results} presents and discusses the numerical results. Sections ~\ref{ss:result-IF-RelatIF} and \ref{ss:result-anomaly-detection} concern tools that increase interpretability of the ML model, i.e., influence functions and relative influence functions (RelatIFs). Sections~\ref{ss:result-LEES} and~\ref{ss:result-RUE} focus on increasing the reliability of the model by assessing the extrapolation score (LEs) and uncertainty of the model predictions (RUE).
Section \ref{s:conclusions} summarizes our paper and presents future possible applications and extensions.\\

\begin{figure}[t]
\begin{center}
\includegraphics[width=\columnwidth]{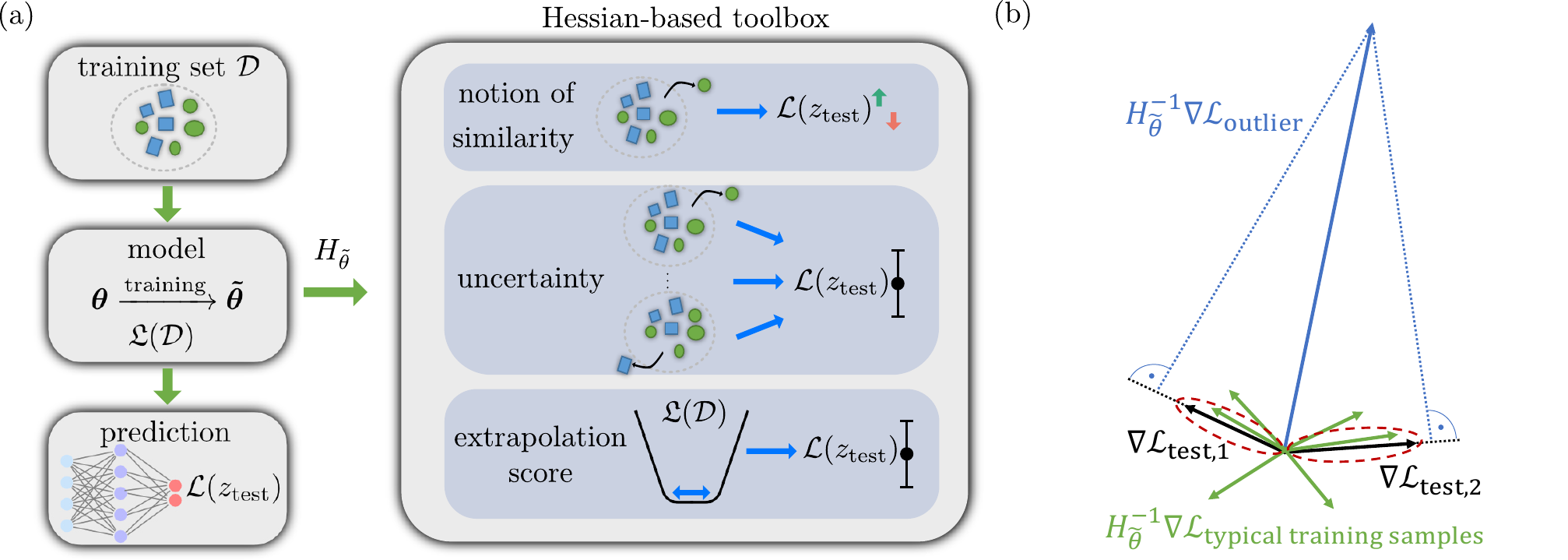}
\end{center}
\caption{(a) The scheme of this study's scope. The ML problem starts with a model depending on parameters $\btheta$. Training a model consists in finding optimal parameters $\btildetheta$ which minimize a training loss function, $\mathcal{L}(\mathcal{D})$, calculated for the training data set, $\mathcal{D}$. The Hessian of the training loss at the minimum, $H_{\btildetheta}$, describes the curvature around the minimum and is a basis for four methods which provide the notion of similarity (influence and relative influence functions), estimation of uncertainty (Resampling Uncertainty Estimation) and extrapolation score (Local Ensembles) of the model prediction. They give insight into the reliability and interpretability of the model after its training. (b) All four methods address the change in the model's predictions due to various actions. This change can be approximated by analyzing the projection of the gradient of the test loss (black arrows) and the gradients of training points (green arrows) corrected for the local curvature described by the Hessian. An outlier (blue arrow) is here a training point being very different from an average one in a data set. Blue dashed lines are large projections of the outlier gradient corrected by the local curvature onto the gradients of the test points.}
\label{fig:intro}
\end{figure}

\section{Methods}
\label{s:methods}

\subsection{Supervised learning}
\label{ss:supervised_learning}

Throughout this work, we focus on supervised ML classification problems with labeled training data $\mathcal{D} = \{z_i\}_{i=0}^n$, with $z_i = (\mathbf{x}_i, y_i)$ where $n$ is the size of the training set.
The input data comes from an input space $\mathbf{x}_i \in \mathcal{X}$, and the labels are $y_i \in \lbrace 0, \ldots, c-1 \rbrace$, where $c \in \mathbb{N}$ is the number of classes in a given problem.
In our setup, the inputs $\mathbf{x}_i$ are the state vectors of a given physical system, and $y_i$ are the corresponding phase labels.

A model, $f$ (in our case a CNN), is determined by the set of $M$ parameters $\btheta = \lbrace \theta_0, \ldots, \theta_{M-1} \rbrace$. The number of parameters, $M$, in our case is of an order of thousand. 
For a given input $\mathbf{x}$, the model outputs a real-valued $c$-dimensional vector, $\mathbf{f}_\mathbf{x} = f(\mathbf{x} ; \btheta)$. The output encodes the prediction of the model, $y' = \mathrm{argmax}(\mathbf{f}_\mathbf{x})$. 
E.g., for a two-class problem, $\mathbf{f}_\mathbf{x}$ could be $[0.1, 0.9]$, which would correspond to predicting a label $y'=1$. Elements of the output vector $\mathbf{f}_\mathbf{x}$ tend to be connected to probabilities of the input belonging to the corresponding classes. However, this interpretation can be misleading in the presence of data set shift \cite{Quinonero09book, Ovadia2019} or non-uniformity of errors \cite{Nushi2018}. Finally, the training process consists in searching the parameter space for the optimal parameters $\btildetheta_{\mathcal{D}} \equiv \btildetheta$ of the ML model, which minimize the empirical risk $\mathfrak{L}(\mathcal{D}, \btheta) = \frac{1}{n} \sum_{z \in \mathcal{D}} \mathcal{L}(z, \btheta)$, also called the training loss, where $\mathcal{L}$ is the loss function.\\

\subsection{Hessian and curvature}
\label{ss:hessian}
The four interpretability and reliability methods discussed in this work are all based on local perturbations of the loss function. They study how a particular action, e.g., removal of a training point, changes the model parameters. This change, in turn, impacts the prediction of the model at a test point, $y'_{\mathrm{test}}$.
The change of the model parameters caused by some action can be approximated using the Hessian matrix of the empirical risk (training loss) calculated at the minimum of the loss landscape at $\btildetheta$, namely $(H_{\btildetheta})_{ij} = \partial^2_{\theta_i \theta_j} \mathfrak{L}(\mathcal{D}, \btheta) |_{\btheta=\btildetheta}$.
$H_{\btildetheta}$ describes the local curvature around the minimum point reached within the training. The eigenvectors of $H_{\btildetheta}$ corresponding to the largest positive eigenvalues indicate directions with the steepest ascent around the minimum. 
The high curvature implies that the training data strongly determines the model parameters along that direction\footnote{If the model parameters are varied along directions of the steepest ascent around the minimum, i.e., along the eigenvectors corresponding to the largest positive eigenvalues of the Hessian, the value of the training loss function changes the most. In other words, these directions are bounded the most by the training data. There are also two empirical observations supporting this claim. Firstly, numerical simulations on the example of a one-dimensional nonlinear regression problem show that gradients of training examples lie along the directions of the highest curvature~\cite{Schulam2020}. Secondly, the analysis of the Hessian spectrum shows that in deep learning problems number of directions with a significant ascent around the minimum is equal roughly to the number of classes in the problem minus one \cite{Sagun2016, Sagun2018, Ghorbani2019}.}.
In contrary to the common intuition, the training of a ML model leads to a local minimum or a saddle point \cite{Dauphin14, Sagun2016, Alain19}: the vast majority of the eigenvalues is close to zero, indicating various flat directions and some small negative eigenvalues are also present, indicating directions with negative curvature. Such a non-positive curvature around the minimum, in general, does not affect the quality of the model predictions but may cause problems when working with the Hessian.\\

All the methods we study in this work approximate how the change of parameters impacts the model predictions, but the reason for the change of parameters is different for each method. Influence functions and RelatIF study the removal of a single training point from a training data set, RUE analyzes training on various samples of the training data set, while LEs modify model parameters in the flat directions of the Hessian. These methods aim to answer different questions regarding the reliability and interpretability of the model, and we will discuss them in detail in the following sections.

\subsection{Leave-one-out training, influence function, similarity, and relative influence function}
\label{ss:IF_RelatIF}
\textbf{Leave-one-out training.} Let us consider a model trained on $n$ training points and making a prediction at a test point.
Now, we remove a single training point $z_r$ from the training set $\mathcal{D}$, $\mathcal{D} \rightarrow \mathcal{D}_{\setminus z_r}$, retrain the model, and check the influence of this removal on the test loss. 
If the prediction is now worse (resp. better), i.e., the test loss is higher (resp. lower), then $z_r$ is a helpful (resp. harmful) training example for this specific test point. If the prediction stays the same, $z_r$ is not influential to this prediction. With such an analysis, called leave-one-out (LOO) training, we can therefore judge how influential a certain training point is for a test prediction.\\

\textbf{Influence functions.} Retraining the model is, however, expensive, and an approximation of the LOO training was proposed and named influence functions~\cite{Cook77, Cook80, Cook82}. It was then ported to ML applications by Koh \& Liang \cite{Koh17, Koh19}. The influence function reads
\begin{equation}
\label{eq:IF}
\mathcal{I}(z_\text{r}, z_{\text{test}}) =  \frac{1}{n} \nabla_{\btheta} \mathcal{L}(z_{\text{test}}, \btildetheta)^T H^{-1}_{\btildetheta} \nabla_{\btheta} \mathcal{L}(z_\text{r}, \btildetheta) \equiv \frac{1}{n} \nabla \mathcal{L}_{\mathrm{test}}^T H^{-1}_{\btildetheta} \nabla \mathcal{L}_{\mathrm{r}}  \,,
\end{equation}\\
and it estimates the change of the test loss for a chosen test point $z_{\text{test}}$ after the removal of a chosen training point $z_{\text{r}}$. $\nabla_{\btheta} \mathcal{L}(z_\text{test}, \btildetheta)$ is the gradient of the loss function of the single test point, and $\nabla_{\btheta} \mathcal{L}(z_\text{r}, \btildetheta)$ is the gradient of the loss function of the single training point whose removal's impact is being approximated. Both are calculated at the minimum $\btildetheta$ of the training loss landscape.\\

\textbf{Geometrical interpretation.} The influence function \eqref{eq:IF} can be written as the inner product of $- \nabla \mathcal{L}_{\mathrm{test}}$ and $-H^{-1}_{\btildetheta} \nabla \mathcal{L}_{\mathrm{r}}$~\cite{Barshan2020}, where the term $- H^{-1}_{\btildetheta} \nabla \mathcal{L}_{\mathrm{r}}$ describes an approximated change in parameters $\btildetheta \rightarrow \btheta'$ due to the removal of the training point $z_{\text{r}}$ (for a derivation, see appendix A of Ref.~\cite{Koh17}). This formulation emphasizes the geometric interpretation of influence functions, which is a projection of the approximated change in parameters due to the removal of a training point onto the test sample's negative loss gradient [see figure~\ref{fig:intro}(b)]. The term $- H^{-1}_{\btildetheta} \nabla \mathcal{L}_{\mathrm{r}}$ can also be understood as a Newton step \cite{Goodfellow2016book} towards a new minimum resulting from a removal of $z_{\text{r}}$. Note that the same term involves scaling by the inverse of eigenvalues of $H^{-1}_{\btildetheta}$. In other words, we see that the influence function is a scalar product of the gradients $\nabla \mathcal{L}_{\mathrm{test}}$ and $\nabla \mathcal{L}_{\mathrm{r}}$ accounting for a local curvature of the loss landscape described by $H_{\btildetheta}$. The resulting value of influence functions depends on two factors: how similar are the test and the removed training point and how representative they are in the data set.\\

\textbf{Similarity measure.} Firstly, the more similar the test point and the removed training point are, the larger is the value of the influence function between them. More specifically, the largest influence is for the change in parameters which is along the direction of $\nabla \mathcal{L}_{\mathrm{test}}$. It happens when the gradients $\nabla \mathcal{L}_{\mathrm{test}}$ and $\nabla \mathcal{L}_{\mathrm{r}}$ are aligned in the parameter space, corrected for the local curvature of the loss landscape, so when the test point, $z_{\text{test}}$, is similar to the removed training point, $z_{\text{r}}$. Note that by similarity here we understand the distance in the model's internal representation, so in the model's parameter space, corrected by the local curvature described by the Hessian. This similarity is different than, e.g., similarity as a distance of input vectors $z_{\text{r}}$ and $z_{\text{test}}$ in the input space $\mathcal{X}$ or the similarity in the Euclidean parameter space. In particular, the predictive model and especially neural networks can be highly nonlinear and may use an internal representation in which similar (close) points are far away in both the input and the Euclidean parameter space. We can then define a model's similarity measure between data points $z_i$ and $z_j$ equal to~\cite{Schulam2020}
\begin{equation}
\label{eq:similarity}
S (z_i, z_j) = [\nabla_{\btheta} \mathcal{L}(z_i, \btildetheta)^T H^{-1}_{\btildetheta} \nabla_{\btheta} \mathcal{L}(z_j, \btildetheta)]^2  \equiv [\nabla \mathcal{L}_i^T H^{-1}_{\btildetheta} \nabla \mathcal{L}_j]^2 \propto \mathcal{I}(z_i, z_j)^2\,.
\end{equation}

\textbf{Representative data and outliers.} The second factor impacting the value of the influence function (and therefore the similarity measure) is the direction along which $\nabla \mathcal{L}_{\mathrm{test}}$ or $\nabla \mathcal{L}_i$ lie. For example, the gradient may be aligned with the eigenvectors of $H_{\btildetheta}$ corresponding to the largest eigenvalues, which are the directions where the training data strongly determines the model parameters. Such an alignment happens for the most common or representative data points. The gradient also can point in the direction of one of many eigenvectors with almost zero eigenvalues, which may happen for distinct or unrepresentative data points called outliers. Due to projection onto the inverse of $H_{\btildetheta}$ and scaling by the inverse of corresponding eigenvalues, the influence function is larger for gradients pointing in the flat curvature of $H_{\btildetheta}$ than for gradients pointing to the high curvature. Therefore, the values of influence functions between two data points are determined by how similar the two data points are from the model's perspective and how representative these data points are in the data set.\\

\textbf{Sensitivity to outliers.} A careful reader notices that influence functions as well as the similarity measure $S (z_i, z_j)$ may then be sensitive to outliers, i.e., data points with extreme values that significantly deviate from the majority of data points \cite{Hendrycks2018}. The removal of such an outlier can cause a large change in parameters. Therefore, the outlier is likely to have a large influence on a wide range of test samples, having a global effect on the test set. This global effect is visualized in figure~\ref{fig:intro}(b), where the blue gradient of the outlier projected onto the inverted Hessian space has large projection lengths with the gradients of two very different test points. Conversely, the removal of a typical training example $z_i$ whose gradient points towards high curvature of the Hessian [green arrows in figure~\ref{fig:intro}(b)] causes a small change in parameters and has a significant influence only for similar test points (circled in red in the figure).\\

\textbf{RelatIF.} Barshan \textit{et al.} \cite{Barshan2020} proposed then a variant of influence functions that takes advantage of the similarity measure but eradicates the influence of unrepresentative data points and outliers. The method is called RelatIF and restricts the pool of influential points to the most similar ones [see the gradients circled in red in figure~\ref{fig:intro}(b)]. Mathematically, it amounts to introducing a normalization to the influence function's formula
\begin{equation}
\label{eq:RelatIF}
\mathcal{I}_{\mathrm{R}}(z_\text{r}, z_{\text{test}}) =  \frac{\mathcal{I}(z_\text{r}, z_{\text{test}})}{|| H^{-1}_{\btildetheta} \nabla_{\btheta} \mathcal{L}(z_\text{r}, \btildetheta) ||} = \frac{\frac{1}{n} \nabla \mathcal{L}_{\mathrm{test}}^T H^{-1}_{\btildetheta} \nabla \mathcal{L}_{\mathrm{r}}}{|| H^{-1}_{\btildetheta} \nabla \mathcal{L}_\text{r} ||}  \,.
\end{equation}\\
Both tools increase the interpretability of the ML model by indicating what the model regards as similar. The influence functions' focus on the unrepresentative data points also allows one to judge the model's reliability by finding outliers in the training data set. The model's reliability is the central issue addressed by two other methods discussed in this work, namely Resampling Uncertainty Estimation and Local Ensembles.

\subsection{Resampling Uncertainty Estimation}
\label{ss:RUE}

The Resampling Uncertainty Estimation (RUE)~\cite{Schulam2020} aims at assessing the uncertainty of predictions of the ML model. It can be applied to the already trained model and requires no specific architecture or learning scheme in contrast to, e.g., Bayesian methods, which are the most common approach in ML for assessing uncertainty \cite{Graves2011, Gal2016}. The RUE method makes use of two important criteria to judge whether a prediction is reliable: the density criterion and local fit criterion \cite{Schulam2020}. The density criterion states that a prediction at the input $z_{\text{test}}$ is reliable if there are samples in the training data that are similar to $z_{\text{test}}$. The local fit criterion states that a prediction at the $z_{\text{test}}$ is reliable if the model has a small error on samples in the training data similar to $z_{\text{test}}$. Both criteria hinge upon a measure of similarity defined in Eq.~\eqref{eq:similarity} and can be addressed with the bootstrap sampling.\\

To quantify uncertainty, the RUE algorithm makes $b$ 'bootstrap' samples created by sampling with replacement from the uniform distribution over the original training data set. Let us start from the original data set $\mathcal{D}$ containing each training data point once, indicated with the short-hand notation by $\mathcal{D} [1,1, \ldots, 1]$. We can then create a bootstrap sample by drawing the same point more than once and omit others, e.g., $\mathcal{D}_b [2,0,3 \ldots, 1,0]$, which stands for taking twice the first training example, omitting the second training example, etc. If a ML model trained on $\mathcal{D} [1,1, \ldots, 1]$ converges to parameters $\btildetheta$, $b$ ML models trained on $b$ different bootstrap samples converge to similar model parameters $\btheta_b'$. Now we can make $b$ predictions at the same test point $z_\mathrm{test}$ with $b$ ML models and calculate the variance of the test loss across these $b$ models. Small variance means that a prediction of the original model can be trusted, while large variance means that it is not reliable. Intuitively, this variance estimates how much the model prediction would change if we fitted the model on different data sets drawn from the same distribution as the original training data. This intuition is connected to the idea of the classical bootstrap~\cite{Efron1986}. \\

As for the LOO training, such a retraining procedure is prohibitively expensive. Therefore, one can make a similar approximation of the change of the model parameters due to the removal of some training examples and adding copies of others to the training set within a bootstrap sample. We can approximate the new parameters via
\begin{equation}
\label{eq:RUE}
\btheta_b' \approx \btildetheta - H^{-1}_{\btildetheta} \cdot L \cdot w_{\Delta_b} \,,
\end{equation}
where $w_{\Delta_b}$ is a vector of differences between the composition of the original $\mathcal{D} [1,1, \ldots, 1]$ and $\mathcal{D}_b$. E.g., for $\mathcal{D}_b [2,0,3, \ldots, 1,0]$, $w_{\Delta_b}$ is $[1,-1,2,\ldots,0, 1]$. $L$ is a matrix of all the $n$ single training loss gradients w.r.t. every model parameter, so it has the size $M$ x $n$, and it takes the form $L = [\nabla_{\btheta} \mathcal{L}(z_0, \btildetheta),\hdots,\nabla_{\btheta} \mathcal{L}(z_{n-1}, \btildetheta)]^T$.
In the next step, one generates predictions at a test point with $b$ models with approximated parameters $\btildetheta_b$, obtaining $b$ test losses based on $\mathbf{f}_b = f(\btildetheta_b, x_\mathrm{test})$.
Finally, we calculate the variance of $b$ test losses, i.e., the average of the squared deviations from the original test loss.\\

\subsection{Extrapolation Score with Local Ensembles}
\label{ss:LEES}

We say the prediction at a test point is underdetermined if many different predictions are equally consistent with the constraints posed by the training data and the learning problem specification (i.e., the model architecture and the loss function). An example of such behavior is when a model trained on the same training data arrives at different predictions depending on the choice of a random seed and, therefore, relies on arbitrary choices outside the learning problem specification. Intuitively, underdetermination can be understood as the model converging during the optimization process to various distinct points in a flat basin forming a minimum. As discussed in section~\ref{ss:hessian}, the training data puts limited constraints on the flat directions around the minimum. However, changing the model parameters in these directions still can impact predictions at test points drawn from a different distribution than the training set, i.e., so-called out-of-distribution (OOD) test points~\cite{Teney2020}. A reliable model should warn the user when making a prediction at such a test point.\\

Local Ensembles (LEs) are a method to detect the underdetermination at test time in a pre-trained model \cite{Madras2020}. LE consists of local perturbations of the parameters of the trained model that fit the training data equally well, i.e., have the same value of training loss. In other words, we perturb the parameters of the model only in the directions of the Hessian eigenvectors corresponding to close to zero eigenvalues, meaning we explore only flat basins around the minimum. Analogously to RUE, the next step is to make predictions at the same test point $z_\mathrm{test}$ with LE models and calculate the variance of the test loss within the LE. Madras \textit{et al.} \cite{Madras2020} found an even simpler approximation of this variance for a test point, $z_\mathrm{test}$, and named it a LE extrapolation score
\begin{equation}
\label{eq:LEES}
\mathcal{E}_m (z_\mathrm{test}) = || U_m^\top \nabla_{\btheta} \mathcal{L} (z_\mathrm{test}, \btildetheta) ||_2 = || U_{m}^\top \nabla \mathcal{L}_\mathrm{test} ||_2 \,.
\end{equation}
$U_m$ is a matrix of $(M - m)$ Hessian eigenvectors spanning a subspace of low curvature, i.e., after removing $m$ eigenvectors corresponding to largest eigenvalues and, therefore, to directions with the highest curvature, which are well-constrained by training data. The authors of the method admit that choosing $m$ is not a trivial task, with the danger of omitting under-constraint directions if $m$ is set too high or including well-constraint directions if $m$ is set too low. In this work, we choose the smallest possible $m$ for which $\mathcal{E}_m$ starts to converge for all test points ($m=12$ in figure~\ref{fig:LEES}).

\subsection{Practical aspects of the Hessian computation}
\label{ss:numerical_challenges}
A careful reader may have noticed two numerical challenges resulting from Eqs.~\eqref{eq:IF}-\eqref{eq:LEES}. Firstly, the calculation of influence functions, RelatIF, and RUE requires inverting the Hessian of a training loss which in deep learning is known to be highly non-convex \cite{Choromanska2015}. As we pointed out in section~\ref{ss:hessian}, the optimization usually leads to a critical point with a majority of flat or almost flat directions (corresponding to eigenvectors with zero or close to zero eigenvalues) and a small number of directions of negative curvature (corresponding to eigenvectors with negative eigenvalues). The inverse of a matrix exists only if it is positive-definite (has only positive eigenvalues). Therefore, Koh \& Liang proposed to add a damping term to the Hessian \cite{Koh17}, $\lambda \, I $, with $I$ being the identity matrix and $\lambda$ being larger than the absolute value of the largest negative Hessian eigenvalue, $|E_0|$. It is equivalent to $L_2$ regularization \cite{Barshan2020} and amounts to shifting all eigenvalues by $\lambda$, guaranteeing the existence of the Hessian inverse. Regardless of the exact value of the damping, the Hessian-based toolbox keeps giving meaningful results \cite{Koh17}. We usually choose $\lambda \approx |E_0| + 0.01$, except for RUE where the authors explicitly state that the smallest eigenvalue of the damped Hessian needs to be around one, rendering $\lambda_{\mathrm{RUE}} = \lambda + 1$.\\

Secondly, the calculation of the Hessian matrix for a model with a large number of parameters can be very demanding. Fortunately, we do not need to calculate the full Hessian, only Hessian-vector products [e.g., $H^{-1}_{\btildetheta} \nabla \mathcal{L}_{\mathrm{r}}$ in Eqs. \eqref{eq:IF} and \eqref{eq:RelatIF}] or the top part of the Hessian spectrum, which significantly reduces the computational complexity of the problem \cite{Agarwal2017}. The inverse of the Hessian for influence functions, RUE, and RelatIF can be approximated with so-called stochastic approximation with LiSSA \cite{Agarwal2017}. Additionally, the authors of RelatIF approximated the normalization factor in Eq.~\eqref{eq:RelatIF} with a method related to K-FAC \cite{Martens2015}. On the other hand, LE needs no inverse but an ensemble subspace of eigenvectors with zero or close to zero eigenvalues. The authors of LE proposed to use the Lanczos iteration \cite{Lanczos1950} to calculate $m$ eigenvectors with the largest eigenvalues, build an $m$-dimensional subspace (of highest curvatures), and create its orthogonal complement, namely the ensemble subspace (of flat directions). There is also a Python library called PYHESSIAN designed to tackle Hessian-based problems \cite{Yao2020}. Within this paper, however, we calculate the Hessian explicitly, due to the limited size of our neural network.

\section{Results}
\label{s:results}
In this section, we present how the interpretability and reliability methods described in sections \ref{ss:IF_RelatIF}-\ref{ss:LEES} extract additional information from a trained model allowing for a better understanding of its predictions and the physics underlying the learned problem.\\

\begin{figure}[t]
\begin{center}
\includegraphics[width=0.7\columnwidth]{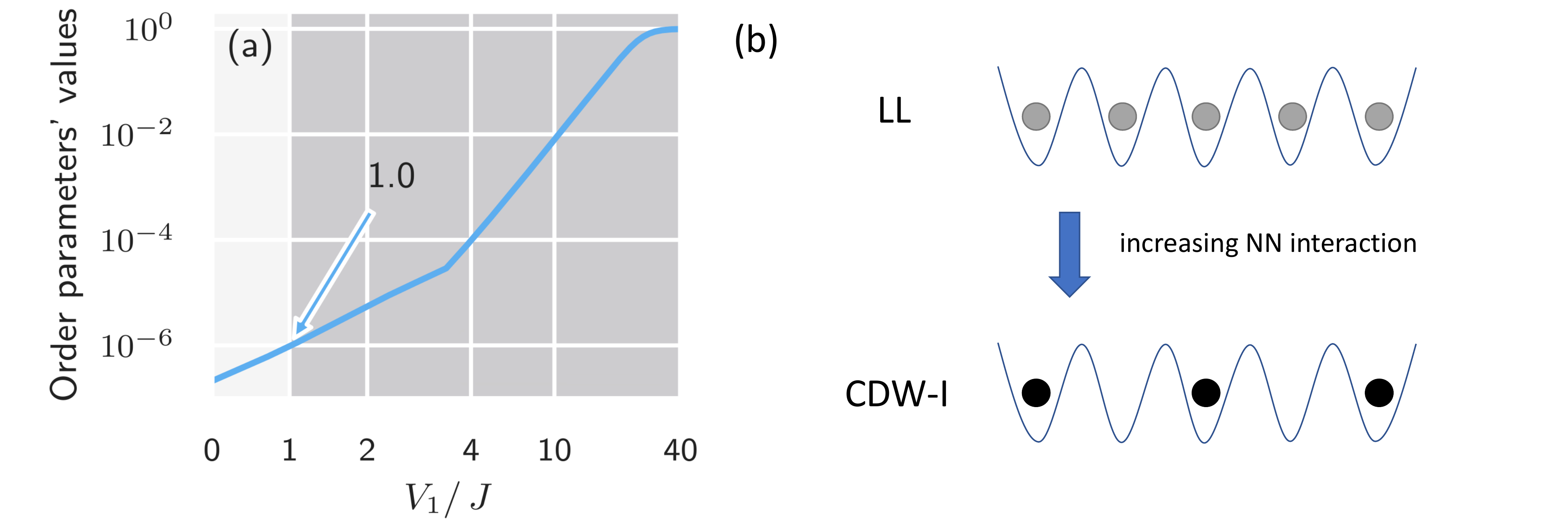}
\end{center}
\caption{The transition line learned by a ML model. (a) Order parameter, $O_{\mathrm{CDW-I}}$, governing the transition is the average difference between the nearest-site densities. It is zero in the Luttinger liquid (LL) phase, and grows to 1 in the charge-density wave (CDW-I) phase. (b) Schematic representation of two phases. The LL phase is uniform, while the CDW-I phase is characterized by a density pattern 1010.}
\label{fig:transition_line}
\end{figure}

\textbf{Physical model.} We show the functionalities of the four interpretability and reliability methods on the example of a CNN trained to recognize phases of the spinless half-filled one-dimensional (1D) Fermi-Hubbard model \cite{Dutta15}. This model describes fermions hopping between neighboring sites with amplitudes $ J $ (set to 1 throughout the paper) and interacting with nearest neighbors with strength $V_1$,
\begin{equation}
\label{eq:ham}
\hat{H} = - J \sum_{\la i,j \ra} c_i^{\dagger} c_j + V_1 \sum_{\la i,j \ra} n_i n_j \,,
\end{equation}
where $c_i^{\dagger}$ and $c_i$ are fermionic creation and annihilation operators at site $i$, respectively, and $n_i = c_i^{\dagger} c_i$ is the number operator. We calculate eigenvectors with exact diagonalization using QuSpin and SciPy packages~\cite{Weinberg17, SciPy}. The detailed description of the data set can be found in our previous work~\cite{Dawid20NJP}. We focus on a single transition line resulting from a competition between hopping and nearest-neighbor interaction, presented in figure~\ref{fig:transition_line}. It starts in the gapless Luttinger liquid (LL) phase, and the increase of $V_1$ leads to a phase transition to a gapped charge-density wave (CDW-I) with density patterns 1010~\cite{Hallberg90, Mishra11}. The order parameter describing the transition, $O_{\mathrm{CDW-I}}$, is the average difference between the nearest-neighbour densities. It is zero in the LL phase and grows to 1 in the CDW-I phase.\\

\textbf{Similarity learned by a ML model.} We feed the CNN with ground states expressed in the Fock basis, labeled with their appropriate phases, calculated for a 12-site or 14-site system. The architectures of the used CNNs are presented in Appendix~\ref{app:architectures}. Intuitively, we could expect that the most similar quantum states, according to the ML model, are those generated for the most similar $V_1$. Instead, as we showed in Ref.~\cite{Dawid20NJP}, the similarity learned by the ML model is based on the order parameter (or something related to it), as it is a much better discriminator between the phases. Therefore, a well-trained model sees all LL data points as very similar (as they all have a zero order parameter), while the similarity of CDW-I data points depends on $V_1$ (on this side, the order parameter continuously goes up from 0 to 1).

\subsection{Comparison between influence and relative influence functions}
\label{ss:result-IF-RelatIF}

\begin{figure}[t]
\begin{center}
\includegraphics[width=0.9\columnwidth]{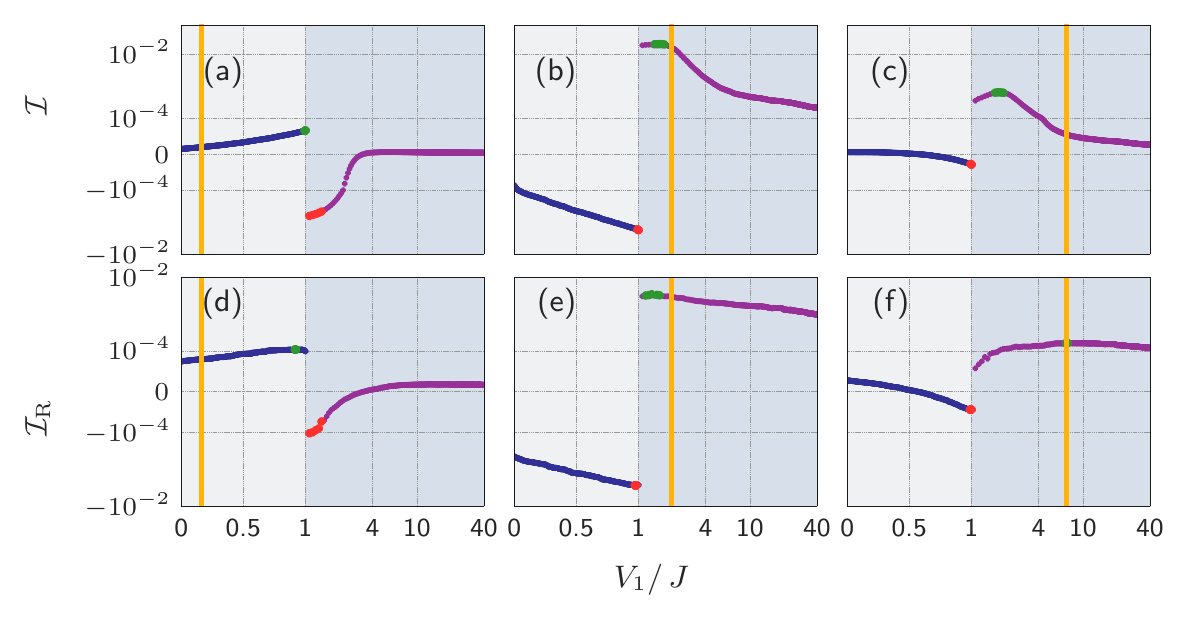}
\end{center}
\caption{Influence functions ($\mathcal{I}$) vs. RelatIFs ($\mathcal{I}_{\mathrm{R}}$). (a)-(c) Influence functions and (d)-(f) RelatIFs between the training set and the single test point indicated by the orange vertical line. Every column analyzes the same test point. We mark the five most helpful (harmful) training points with green (red) color. Blue (purple) training points belong to the LL (CDW-I) phase. We mark the transition point with the change of the background color.}
\label{fig:IFvsRelatIF}
\end{figure}

While in Ref.~\cite{Dawid20NJP} we studied in detail the potential of influence functions for interpretability, we here focus on the differences between influence functions and relative influence functions (RelatIFs). To this end, we train a CNN on the eigenvectors of the 12-site 1D Fermi-Hubbard model with the labels indicating phases they belong to, i.e., LL or CDW-I. We then analyze the trained model with influence functions [Eq.~\eqref{eq:IF}] and RelatIFs [Eq.~\eqref{eq:RelatIF}] and present results in figure~\ref{fig:IFvsRelatIF}. Each column of figure~\ref{fig:IFvsRelatIF} presents both methods calculated for the same test points, indicated by the orange vertical lines. In the case of influence functions (first row of figure~\ref{fig:IFvsRelatIF}), the most helpful training points are the most similar to the test point and the most unrepresentative in the data set. RelatIFs, presented in the second row of figure~\ref{fig:IFvsRelatIF}, are expected instead to ignore how representative data is and to indicate the most similar training points to the test point, paying less attention to outliers.\\

\textbf{Behavior in the CDW-I phase.} The usefulness of RelatIF can be seen in the last column of figure~\ref{fig:IFvsRelatIF} when comparing influence functions and RelatIFs calculated for the whole training set and the test point located deeply in the CDW-I phase [panels (c) and (f), respectively]. According to influence functions, the location of the most helpful points balances between those being the most unrepresentative (close to the phase transition) and most similar (with the most similar order parameter). As a result, the most helpful points do not follow the test point into the deep region of the CDW-I phase but get stuck instead. RelatIF ignores how representative data is, and the five most helpful training points follow the test point much deeper to the CDW-I phase because they have the most similar order parameter. Let us now compare the results for the test point in the transition regime in panels (b) and (e). Influence functions and RelatIFs yield here nearly the same results. The test point is in the unrepresentative regime, so the RelatIF's correction is not needed to observe which data is, in reality, most similar. Still, values of RelatIFs for CDW-I points are much closer to each other than values of influence functions due to a smaller focus on distinctive points close to the phase transition.

\textbf{Model-agnostic RelatIFs.} A careful reader who is familiar with our previous work \cite{Dawid20NJP} may notice that in this previous work, influence functions for test points located deeply in the CDW-I phase have a different pattern than in figure~\ref{fig:IFvsRelatIF}(c). We discuss this observation in Appendix~\ref{app:architectures}. The key point is that different influence patterns may, in particular, be caused by a different level of the ML model's focus on outliers. In this sense, RelatIFs, which apply a correction for the model's focus on unrepresentative data, are more model-agnostic.\\

\textbf{Behavior in the LL phase.} Finally, let us compare panels (a) and (d) of figure~\ref{fig:IFvsRelatIF} with the test point located deeply in the LL phase. 
RelatIFs' values of the LL training points in panel (d) form a more flat line (are less varied) than corresponding influence values in panel (a). In other words, RelatIFs indicate that the LL points are more similar to each other, which agrees with the zero order parameter of the LL. 
Now let us compare the most influential training points between the methods. The five most helpful training points (marked in green), according to influence functions in panel (a), are LL data points closest to the transition, while the five most harmful training points (marked in red) are CDW-I data points closest to the transition. They are the most similar to the test point but labeled oppositely, so they confuse the model. RelatIFs indicate the same training points as the most harmful, as the logic behind it is independent of the data representativeness. However, the most helpful points in panel (d) are shifted deeper towards the LL phase than in panel (a). With unrepresentative data being less important, this is the desired direction, i.e., taking the most helpful points further away from the unrepresentative transition regime. The persistent non-zero slope of the LL points may result from the imperfect removal of the impact of unrepresentative data by the RelatIF normalizer and the finite-size effect present in the 12-site Fermi-Hubbard model.\\

Therefore, we can study the similarity with influence functions unless the model focuses predominantly on outliers. However, as we present in the next section, we can use influence functions' property to focus on outliers to our advantage for anomaly detection. When we need to study similarity, RelatIFs provide a needed correction to ignore how representative data points are.

\subsection{Influence functions for anomaly detection}
\label{ss:result-anomaly-detection}

\begin{figure}[t]
\begin{center}
\includegraphics[width=0.8\columnwidth]{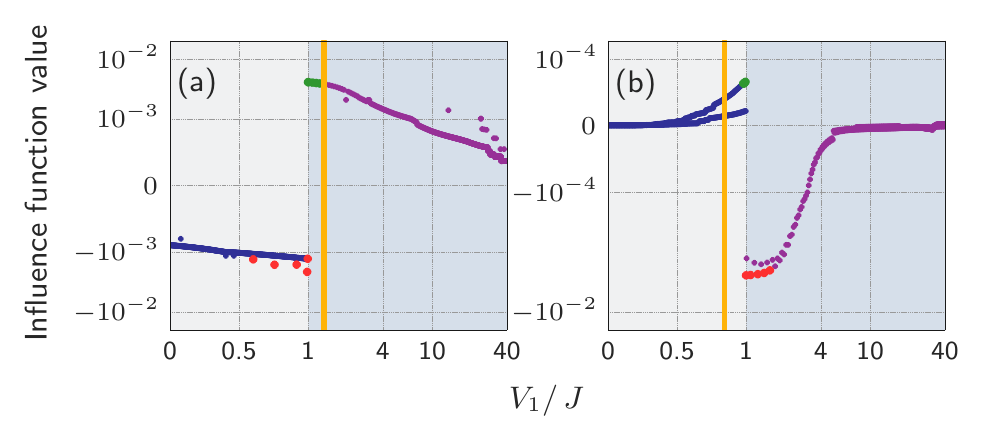}
\end{center}
\caption{Influence functions between the training set and the single test point indicated by the orange vertical line. We mark the five most helpful (harmful) training points with green (red) color. Blue (purple) training points belong to the LL (CDW-I) phase. We indicate the transition point by the change of the background color. (a) Influence functions show anomalies in the training data. Here, in the sign-imbalanced data set, training points diverging from the smooth lines are negative-sign points. (b) In the sign-balanced data set, influence functions form two subgroups. The first one is created by the training points with the same global sign as the negative-sign test point, the other - with the opposite sign. The ML model is not truly invariant with respect to this property, even if the classification accuracy is high. Note the use of the symmetric-log scale.}
\label{fig:anomaly_detection}
\end{figure}

We continue working with the data set of the eigenvectors of the 12-site 1D Fermi-Hubbard model with labels. The global sign of such an eigenvector is not a physical observable, and therefore a well-generalizing model may ignore this property. To challenge this intuition, we prepare two data sets differing only in the distribution of the global sign. The first data set is composed of a large majority of positive-sign eigenvectors and several negative-sign eigenvectors. In the second, half of the eigenvectors have a positive global sign and the other half - a negative global sign. We call them the 'sign-imbalanced' and 'sign-balanced' data sets, respectively.\\

\textbf{Detection of outliers.} The CNN trained on the sign-imbalanced data set has high accuracy on positive-sign test points. Regarding negative-sign test points, the CNN correctly classifies them on the CDW-I side but has lower accuracy on the LL side. This suggests that the ML model grasped the global-sign invariance only to a limited level. Figure~\ref{fig:anomaly_detection}(a) presents the influence functions between the training data and the single positive-sign test point near the transition (marked with the orange vertical line). The pattern is dominated by two smooth lines on both sides of the phase transition, formed by positive-sign training points, as in figure~\ref{fig:IFvsRelatIF}(b) in the previous section.
However, there are several single training points that are outside the continuous patterns. These 'outsiders' are negative-sign training points that the model regards as distinct from positive training points, in the sense of the similarity as described in section~\ref{ss:IF_RelatIF}. Influence functions then immediately pinpoint anomalies in the training data and improve the reliability of the model. The non-zero influence of the outliers and the decent test accuracy on negative-sign points indicate that the model gains some information from a minority of negative-sign training points. However, a model can develop different similarity measures depending on the training process and its architecture. For example, all anomalies (or outliers) can have zero influence regardless of the test point. It shows that the model ignores them during the training (which may be encouraged by large regularization), and we could further use such knowledge to improve the model.\\

\textbf{Global sign (in)variance of the ML model.} We also train a CNN on the sign-balanced data set. The model achieves high accuracy on both positive- and negative-sign test data, suggesting that it learns the global-sign invariance and ignores this property in the decision-making process. This would mean the model is genuinely invariant to the global sign. We check this interpretation with influence functions plotted in figure~\ref{fig:anomaly_detection}(b) and calculated for the whole sign-balanced training set and the single negative-sign test point marked again with the orange line. If a model is sign-invariant, we should reproduce the smooth patterns of figure~\ref{fig:IFvsRelatIF}(a)-(c). Surprisingly, the training points form two subgroups of influence, following their global sign. The subset of training points with the same global sign as the test point reproduces the smooth shape of figure~\ref{fig:IFvsRelatIF}(a)-(c), while the subgroup with opposite sign separates. The separation is the strongest near the transition. In the end, the most influential training points are always the ones with the same global sign. Thus, if a feature, to the best of our knowledge, is irrelevant for the classification, the model may still note the property. Again, this behavior and developed concept of similarity depend on the architecture and training process. Still, in our numerical experiments, we always arrived at a model that recognized the training points' global sign.\\

\textbf{Strategies to make a ML model invariant.} Therefore, we see that having a ML model that is truly invariant to some properties may be a challenging, yet highly rewarding task. A convincing example is Ref.~\cite{Kaming2021}, studying experimental Floquet data, where a property called micromotion had no impact on physical phases in the system but was consistently recognized by a ML model, rendering unsupervised phase classification ineffective. In the end, the authors processed the experimental data with a variational autoencoder, effectively removing this property. Another approach would be using domain adaptation neural networks, which could be trained towards ignoring a chosen property \cite{Ganin2016}.\\

The results presented so far concern two methods: influence functions and RelatIFs. Both analyze the relation between a test point and training point and are especially useful for analyzing the training data or the similarity learned by a model. When assessing the reliability of ML model predictions, more appropriate tools are LEs and RUE.

\subsection{Extrapolation Score with Local Ensembles for out-of-distribution test points}
\label{ss:result-LEES}
\begin{figure}[b]
\begin{center}
\includegraphics[width=0.7\columnwidth]{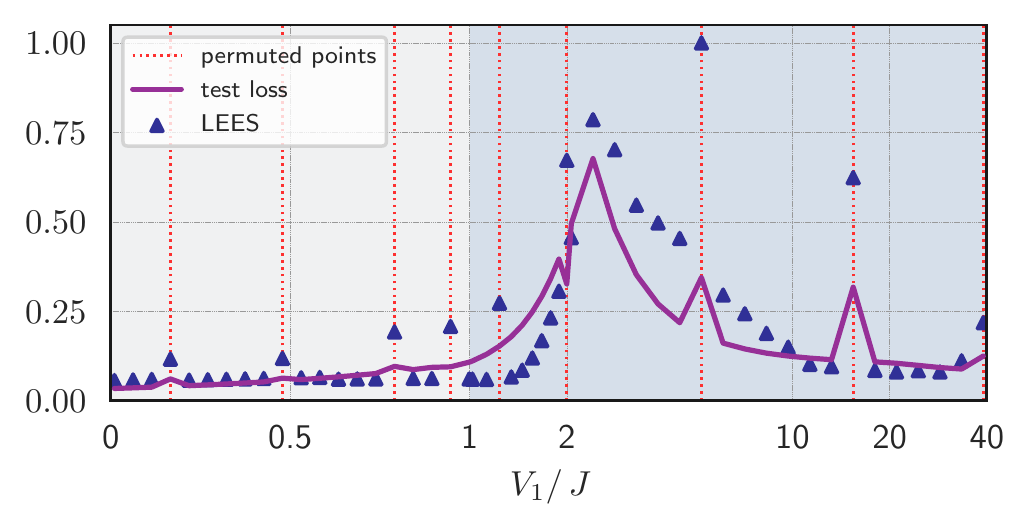}
\end{center}
\caption{Detection of the CNN extrapolation. We plot the minimal test loss (i.e., assuming all predicted labels are correct) with a purple line. We mark the LE-based extrapolation score (LEES) with blue triangles. Red vertical lines symbolize OOD test points. All OOD test points have a higher extrapolation score, but not always they are detected by a minimal test loss. We mark the transition point with the change of the background color. Note the symmetric log scale in the $x$-axis.}
\label{fig:LEES}
\end{figure}
We now challenge the concept of a test loss as a reliability measure. As we will show, the extrapolation score with Local Ensembles (LEs) highlights better out-of-distribution (OOD) test points. To do so, we introduce in our test set a percentage of test elements whose components are randomly permuted. A well-trained ML model, combined with a reliability method, should be able to inform us of the OOD test points.\\

\textbf{Minimal test loss.} Let us start by looking at the loss function for every test point, plotted in figure~\ref{fig:LEES} with a purple line. We here use the minimal version of the loss function (i.e., assuming all predicted labels are correct, more details in appendix~\ref{app:minimal_test_loss}) to mimic a real-life scenario: we ask a trained ML model to make predictions at test points whose labels we do not know. However, here we have access to the ground-truth labels, and we know the model misclassifies the test points generated for $V_1/\,J \in [1,2.1]$. The model wrongly predicts that the points in this interval belong to the LL phase. We see that the minimum test loss in figure~\ref{fig:LEES} is primarily smooth and reaches a maximum around $V_1/\,J = 2.1$. This is the predicted transition point of the model, i.e., for this test point, the model outputs values corresponding to the LL and CDW-I class, which are very close to each other.\\

\textbf{LE-based Extrapolation Score.} We then calculate the Local-Ensemble-based Extrapolation Score (LEES) of the CNN's predictions at the same test set. We plot LEES values as blue triangles in figure~\ref{fig:LEES}. Let us start with the analysis of LEES for the regular test points, ignoring the OOD test points. A first observation is that LEES mostly follows the test loss, which we expect as by its definition LEES is proportional to the gradient of the test loss. Secondly, LEES approaches zero for the predictions at the test points deep in the LL and CDW-I phases. These examples are well-constrained by the typical training examples, and the model needs no extrapolation to make its predictions. Thirdly, we also see that, while LEES is close to zero for all the LL test points, it is not for the CDW-I side. This lack of symmetry is contrary to the test loss, which is non-zero close to the transition, regardless of the phase. Strictly speaking, it means that the gradients of all LL test points are parallel to the gradients of the most typical LL training examples, corrected for the local curvature of the loss landscape, i.e., parallel to some of the Hessian eigenvectors with the largest eigenvalues, which we removed to build a flat LE. The ML model sees them as very similar, needs no extrapolation to make predictions at them, and changes of parameters within the LE make no difference. CDW-I test points, on the other hand, exhibit larger diversity, and the model reaches the same low extrapolation level of its predictions as in the LL only deeply in the phase. This makes perfect sense if we assume that two typical training examples representing two classes are ones with $O_{\mathrm{CDW-I}} = 0$ and 1, respectively. Then we can explain LEES' large values by combining two facts: firstly, it approximately follows the test loss; secondly, it is larger for test points being far from the representative training example from the appropriate class.\\

\textbf{Fate of the OOD test points.} We now focus on the OOD test points, marked in figure~\ref{fig:LEES} with red vertical lines. On the one hand, if one interprets the test loss as the uncertainty, the test loss should be enough to highlight the OOD test points as the corresponding predictions should be less confident than those of the neighboring test points. Partially, this intuition holds as we see significant jumps in test loss for some of the OOD test points, especially for those deep in the CDW-I phase. However, the jumps corresponding to the OOD points are much less prominent on the LL side and disappear completely when getting closer to the transition regime. In particular, prediction at the OOD test point at $V_1/\,J=2$ is recognized as more confident than on the neighbors, which further challenges the concept of the test loss as the uncertainty measure. On the other hand, the LEES perfectly highlights each of them, being significantly larger for all OOD test points than its neighbors. The CNN needs to extrapolate on these unseen, atypical test points which the LEs is detecting. With this method, we can then have a neural network built and trained without any constraints, which informs us of predictions made with a high extrapolation level. It is crucial for the predictions at OOD test points, which are not detected at all by the test loss, here, e.g., those between  $V_1 / \, J=0.75$ and $2$.\\

\subsection{Resampling Uncertainty Estimation to identify the phase transition region}
\label{ss:result-RUE}
We finally use the Resampling Uncertainty Estimation (RUE) method to analyze the uncertainty of predictions of the ML model and therefore identify phase transitions regions. In particular, it is known that finite size effects play a great role in the analysis of phases and should be taken into account when predicting phase transitions. We, therefore, train two copies of the same CNN architecture on two data sets, so on the eigenvectors of the 12-site and 14-site 1D Fermi-Hubbard model with labels corresponding to the LL or CDW-I phase. We choose a CNN architecture that is invariant to the input size (see appendix \ref{app:architectures}).\\

\begin{figure}[t]
\begin{center}
\includegraphics[width=0.99\columnwidth]{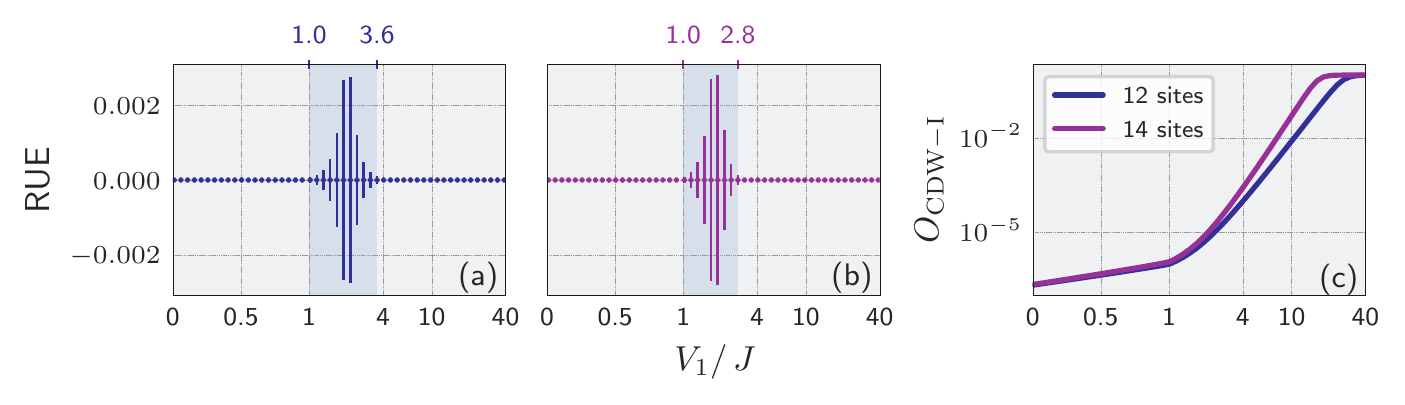}
\end{center}
\caption{RUE vs. the quantum phase transition width. (a)-(b) RUE plotted as error bars of predictions of two CNNs with the same architecture trained on the 12- and 14-site systems. The uncertainty regime is narrower for the 14-site system than for the 12-site one. (c) Order parameters across the same transition for the 12- and 14-site systems. Due to the finite-size effect, the transition is sharper in the 14-site system than the 12-site. }
\label{fig:RUE_12_14_sites}
\end{figure}

\textbf{Visualization of RUE.} To detect the quantum phase transition region, we calculate RUE for the whole test set uniformly spread across the transition line. Panels~\ref{fig:RUE_12_14_sites}(a)-(b) show the results for the CNNs trained on the 12- and 14-site systems, respectively. We represent RUE as error bars as they are the variance of the test loss' change under the different sampling of the original training data\footnote{We could plot RUE as the error bars of the test loss along with the values of the test loss. However, in our case, various sampling of training data set leads to the test loss' change of the order of $|0.001|$, so around 0.1\% of the test loss value, and plotting them together would be infeasible. Secondly, we focus on the width of the regime where the error bars are non-zero, which is better visualized without plotting the test loss values.}. It is important to note that we calculate RUE using the minimal version of the test loss (see the discussion in appendix \ref{app:minimal_test_loss}), assuming that predicted labels are the correct ones. \\

\textbf{Transition regime and error bars.} By construction, we know that the RUE indicates uncertainty caused by the limited number of training examples being similar to the test point or due to the ML model making mistakes on training examples that are similar to the test point. As a result, non-zero error bars cover the whole transition regime where the model has troubles with classification. Note we use here a minimal loss function, so RUE has no information on the ground-truth labels. Nevertheless, the error bars are the largest for the incorrect model predictions (i.e., for $V_1/J$ in the interval $[1,1.6]$ for 12 sites and $[1,1.45]$ for 14 sites). RUE then manages to detect misclassification of test points. Error bars are non-zero also for correct predictions (i.e., in the interval $[1.6,3.6]$ for 12 sites and $[1.45,2.8]$ for 14 sites), but here RUE warns a user that the ML model made these decisions based on the limited number of training data. Moreover, note that our choice of the phase transition point (set to $V_1\,/J = 1$) is to some level arbitrary for numerical reasons discussed in appendix A of Ref.~\cite{Dawid20NJP}.\\

\textbf{Error bars for 12- and 14-site systems.} More importantly, we see that these uncertainty regimes have different widths for two different system sizes (see panels (a) and (b) of figure~\ref{fig:RUE_12_14_sites}). If we set a threshold for RUE's value to $5 \times 10^{-5}$, the uncertainty regime spans between 1 and $3.6\, V_1/\,J$ for 12 sites and 1 and $2.8\, V_1/\,J$ for 14 sites. It is a direct consequence of the finite-size effect because of which the transition is sharper for the 14-site system than for the 12-site one. Figure~\ref{fig:RUE_12_14_sites}(c) depicts the order parameter $O_{\mathrm{CDW-I}}$ for both system sizes, and one can clearly see a sharper phase transition for larger system size. Due to the sharper transition and smaller number of test data with low representation in the training data, the non-zero RUE regime is always narrower in the 14-site case, regardless of the chosen threshold for the RUE's value.\\

Therefore, RUE is a way of providing similarity-based confidence in the ML model predictions.

\section{Conclusions}
\label{s:conclusions}
In this work, we presented four interpretability and reliability methods that are independent of the architecture and the training procedure of the ML models. They rely on the computation of the Hessian of the training loss describing the curvature around the local minimum. We showed how these methods could be applied to ML models that classify many-body physics phase diagrams, here the phase classification of the 1D spinless Fermi-Hubbard model. Our findings are summarized in the following:
\begin{itemize}
\item We compared influence functions and RelatIFs. According to influence functions, the most influential training points were the most similar to the test point and the most unrepresentative in the data set. RelatIF ignored the second aspect and focused on similarity. Here, we mean the similarity as a distance in the model's learned internal representation space. The analysis of this learned similarity, enabled by influence functions and RelatIFs, increases the interpretability of the ML model.

\item Thanks to the focus on unrepresentative data, influence functions immediately pinpointed anomalies in the training set and improved the model's reliability. The model can be better understood when the influence of the training outliers is known. For example, an outlier training point can have zero influence on all the test points. This shows that the model ignores such outliers during the training. In phase-detection tasks, in general, the model should be prone against outliers, and therefore, one can further use such knowledge to improve the model and its training.

\item With the help of influence functions, we also showed that, even if a feature is irrelevant for the phase classification (like a global sign of the wavefunction), the model may still note such features. This finding is consistent with the results found in~\cite{Kaming2021}. These findings challenge our intuition about what really is a feature-invariant model.

\item The test loss calculated by comparing the output of the ML model and the ground-truth label tends to be interpreted as an uncertainty of the ML model. This interpretation is tempting due to the simplicity of the test loss calculation but has limited use and can fail miserably on OOD test points. Therefore, we need other tools to increase our trust in the ML model.

\item We showed how LEs are able to identify predictions that are made by a ML model with a high level of extrapolation. In other words, it shows how sensitive the prediction is to the arbitrary choices outside the learning problem, e.g., the random seed. Thanks to this property, LEES perfectly highlighted all out-of-distribution (OOD) points in our test set.

\item We showed that RUE provided error bars for the ML model's predictions. RUE is large when the training set lacks data similar to the test point or when the ML model makes mistakes on training data similar to the test point. Analysis of RUE values across the test set allowed to assess the phase transition region, which was smaller in the case of the CNN trained on the 14-site system's data than for the 12-site, accordingly with the finite-size effect.

\end{itemize}

The presented functionalities of the four methods do not exhaust the possible applications. For example, influence functions and RelatIFs may be used for building more physics-informed ML models. If we know a proper similarity measure, e.g., based on the order parameter in some solvable regime, we can select a ML model which learned the desired similarity and apply the model later to unknown regimes. Moreover, LEES is helpful for active learning in which the model informs the user which additional data points would be the most informative for the training \cite{Madras2020}. This approach can prove extremely useful for ML based on expensive experimental measurements. LEES and RUE also can be used to detect additional phases in the data by informing the user about the part of test data on which predictions are highly extrapolated or uncertain. The idea is analogous to the use of influence functions in Ref.~\cite{Kaming2021} and to the anomaly detection scheme in Ref.~\cite{Kottmann2020PRL}. Finally, it would be interesting to apply the same toolbox in the context of quantum machine learning with variational quantum circuits, where the Hessian of the loss function can also be computed~\cite{Huembeli21,Mari21}.

\begin{acknowledgments}
We thank Pang Wei Koh and David Madras for useful tips and explanations. We acknowledge David Madras for referring us to the paper on RelatIFs. We also thank Piotr T. Grochowski for fruitful discussions.

An.D. acknowledges the financial support from the National Science Centre, Poland, within the Preludium grant No. 2019/33/N/ST2/03123 and the Etiuda grant No. 2020/36/T/ST2/00588.
M.T. acknowledges the financial support from the Foundation for Polish Science within the First Team programme co-financed by the EU Regional Development Fund and the PL-Grid Infrastructure.
ICFO group acknowledges support from ERC AdG NOQIA, State Research Agency AEI (“Severo Ochoa” Center of Excellence CEX2019-000910-S, Plan National FIDEUA PID2019-106901GB-I00/10.13039 / 501100011033, FPI, QUANTERA MAQS PCI2019-111828-2 / 10.13039/501100011033), Fundació Privada Cellex, Fundació Mir-Puig, Generalitat de Catalunya (AGAUR Grant No. 2017 SGR 1341, CERCA program, QuantumCAT \ U16-011424, co-funded by ERDF Operational Program of Catalonia 2014-2020), EU Horizon 2020 FET-OPEN OPTOLogic (Grant No 899794), and the National Science Centre, Poland (Symfonia Grant No. 2016/20/W/ST4/00314), Marie Sk\l odowska-Curie grant STREDCH No 101029393.
Al.D. acknowledges the financial support from a fellowship granted by la Caixa Foundation (ID 100010434, fellowship code LCF/BQ/PR20/11770012).
\end{acknowledgments}

\subsection*{Data availability}
The source code that support the findings of this study is openly available at \url{https://doi.org/10.5281/zenodo.5148870} \cite{OurRepo}. The physical data can be generated using our open repository \cite{OurRepoNJP}.


------------------------------------------------------------------------------
\appendix
\section{Minimal version of the test loss}
\label{app:minimal_test_loss}

A ML model, $f$, is determined by the set of $M$ parameters $\btheta = \lbrace \theta_0, \ldots, \theta_{M-1} \rbrace$.
For a given input $\mathbf{x}$, the model outputs a real-valued $c$-dimensional vector, $\mathbf{f}_\mathbf{x} = f(\mathbf{x}; \btheta)$. The output encodes the prediction of the model, $y' = \mathrm{argmax}(\mathbf{f}_\mathbf{x})$. 
E.g., for a two-class problem, $\mathbf{f}_\mathbf{x}$ could be $[0.1, 0.9]$ which would correspond to predicting a label $y'=1$.
In a supervised scheme, the loss function, $\mathcal{L} = (\mathbf{x}, \theta)$, compares a model's output, $\mathbf{f}_\mathbf{x}$ with a ground-truth label $y$ of the corresponding input $\mathbf{x}$. 
$\mathcal{L}$ is small when the predicted label $y'$ is the same as the ground-truth label $y$. Moreover, $\mathcal{L}$ gets smaller, the larger are differences between the $y$-element's and other elements' values in the model's output, $\mathbf{f}_\mathbf{x}$. E.g., the $\mathcal{L}$ would be smaller for $\mathbf{f}_\mathbf{x} = [0.1, 0.9]$ than for $[0.45, 0.55]$, even though the predicted label is the same in both cases. For this reason, the elements of the output vector $\mathbf{f}_\mathbf{x}$ tend to be connected to probabilities of the input belonging to the corresponding classes. However, this interpretation can be misleading in the presence of data set shift \cite{Quinonero09book, Ovadia2019} or non-uniformity of errors \cite{Nushi2018}.
Training ends when a minimum of $\mathcal{L}$ is found, and parameters at this minimum are $\btildetheta$.

After training, the model can make a prediction at an unseen test point $y'_{\mathrm{test}} = \mathrm{argmax}(\mathbf{f}_{\mathrm{test}})$, where $\mathbf{f}_{\mathrm{test}} = f(x_{\mathrm{test}}, \btildetheta)$ with a test loss function $\mathcal{L}(x_{\mathrm{test}}, \btildetheta)$.
Within this work, we use two versions of the test loss. The "ground-truth" version is the standard test loss defined in supervised problems and compares the output of the model, $\mathbf{f}_{\mathrm{test}}$ with the ground-truth label $y_{\mathrm{test}}$ of $\mathbf{x}_{\mathrm{test}}$. When the ground-truth label of a test point is unavailable, one can use a "minimal" version of the test loss. It compares the model's output $\mathbf{f}_{\mathrm{test}}$ to the model's predicted label $y'_{\mathrm{test}}$. We stress we use it only during the test stage to imitate the real-life situation when we ask a ML model for predictions at test points we do not know ground-truth labels for.

\section{Architectures of used ML models}
\label{app:architectures}

\begin{figure}[b]
\begin{center}
\includegraphics[width=0.99\columnwidth]{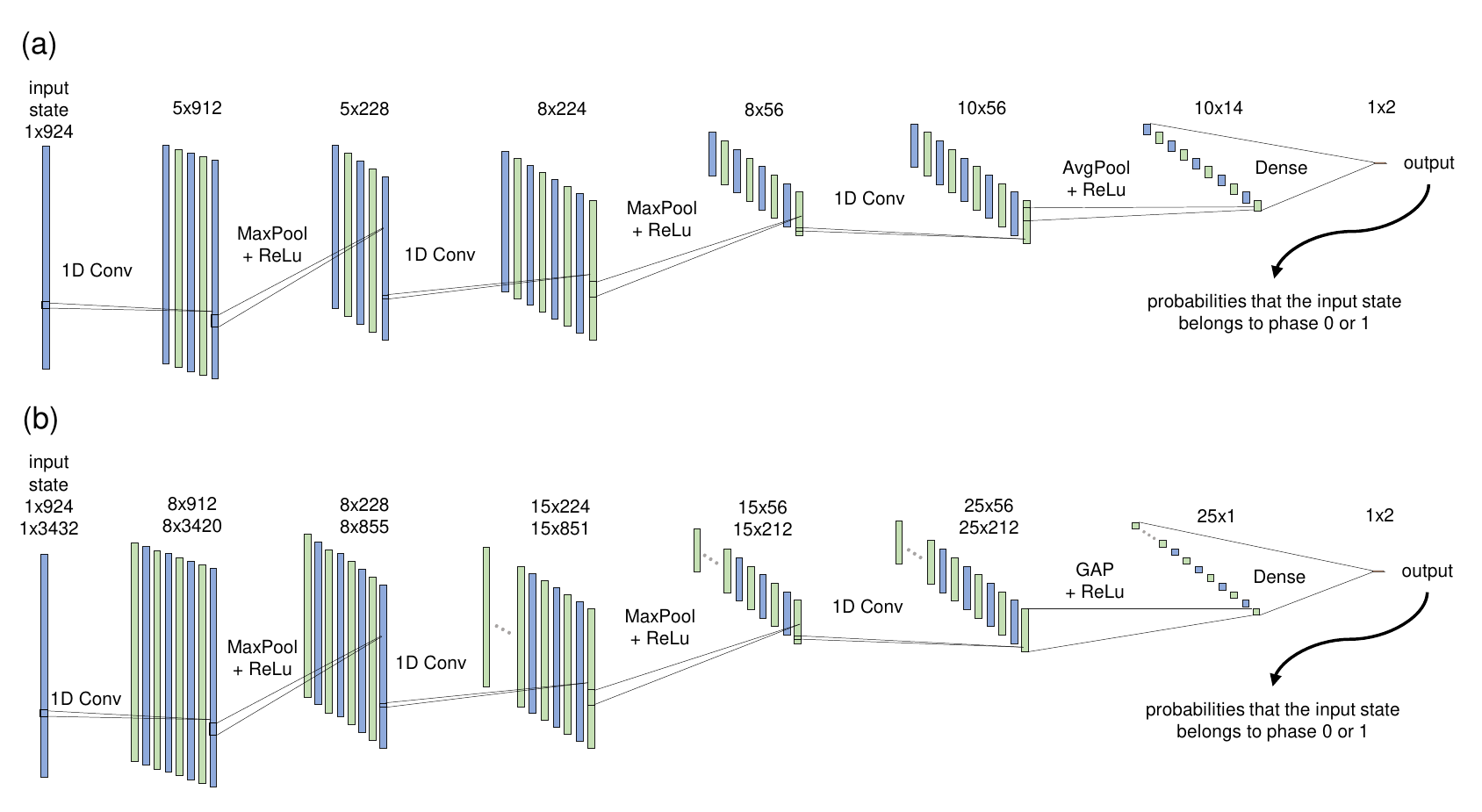}
\end{center}
\caption{Architectures of ML models used in this work. (a) CNN was applied to the classification of eigenstates of the 12-site 1D Fermi-Hubbard model. CNN has 720 parameters. (b) CNN with Global Average Pooling (GAP) is largely invariant to the input size and was applied to the classification of eigenstates of the 12-site and 14-site 1D Fermi-Hubbard model. CNN with GAP has 1675 parameters, regardless of the input size.}
\label{fig:architectures}
\end{figure}

Within this work, we use two architectures for ML models trained to classify phases in the 1D spinless Fermi-Hubbard model. Figure~\ref{fig:architectures} shows them both schematically. We used the CNNs with architecture presented in figure~\ref{fig:architectures}(a) in sections \ref{ss:result-anomaly-detection} and \ref{ss:result-LEES} of this work as well as in our previous paper \cite{Dawid20NJP}. Thanks to its relatively small size (only 720 parameters), its Hessian-based analysis is very efficient. However, it is designed for inputs of size 924, which is the size of eigenvectors of the 12-site Fermi-Hubbard model. 
To have a ML model which is invariant to the input size, we design a CNN architecture with a Global Average Pooling (GAP) layer, which reduces the size of each input filter to one. In figure~\ref{fig:architectures}(b), we list the sizes of convoluted data passing through the model for two input sizes, 924 and 3432, corresponding to 12- and 14-site eigenfunctions of the Fermi-Hubbard model. After the GAP layer, the sizes of convoluted data are the same, which shows how the size-invariance is reached. We use this model in sections \ref{ss:result-IF-RelatIF} and \ref{ss:result-RUE}.

The reason for using the CNN with GAP in section \ref{ss:result-RUE} is simple. We compare there the uncertainty of models trained to classify 12- and 14-site eigenstates, and by choosing the same size-invariant architecture, we minimize the differences between set-ups. We use the CNN with GAP also in section \ref{ss:result-IF-RelatIF} which deserves an additional explanation. In section \ref{ss:result-IF-RelatIF}, we compare the outcomes for influence functions and RelatIFs for the same ML model. When we do it for the CNN from figure~\ref{fig:architectures}(a), the outcomes are very similar. In particular, as presented in Ref.~\cite{Dawid20NJP}, most influential points, according to the influence functions, follow the test point much further into the CDW-I phase, in the same way as RelatIFs. Only the analysis for the CNN with GAP rendered differences between influence functions and RelatIFs described in section \ref{ss:result-IF-RelatIF}. The bottom line is that the similarity measure learned by ML models may depend on their architecture and the hyperparameters determining their training. In particular, models may differ in the magnitude to which they regard the transition points as outliers. In our case, both models (correctly) see them as unrepresentative in the data set, and the influence of the transition points consistently varies from LL and CDW-I points deep in the phases. However, the ML model with GAP (figure~\ref{fig:architectures}(b)) treats transition points as less representative than the smaller ML model without GAP (figure~\ref{fig:architectures}(a)). This is the interpretation of the mathematical fact that the Hessian of the ML model with GAP has larger eigenvalues, i.e., describes a more curved minimum of a GAP model. As a result, representativeness dominates influence, and the normalization brought by RelatIF (Eq.~\eqref{eq:RelatIF}) is necessary to overcome this effect and focus on the similarity.

\bibliographystyle{apsrev4-1_our_style}
\bibliography{BIB_influence_extensive}

\end{document}